# Quantitative comparison of fuel spray images obtained using ultrafast coherent and incoherent double-pulsed illumination

Harsh Purwar*, Kamel Lounnaci, Saïd Idlahcen, Claude Rozé,
Jean-Bernard Blaisot, Thibault Ménard
CORIA UMR-6614, Normandie Université, CNRS, Université et INSA de Rouen
Campus Universitaire du Madrillet, 76800 Saint Etienne du Rouvray, France
*Corresponding author: harsh.purwar@coria.fr

**Abstract**
We present a quantitative comparison between the high-pressure fuel spray images obtained experimentally using classical imaging with coherent and incoherent ultrafast illuminations recorded using a compatible CMOS camera. The ultrafast, incoherent illumination source was extracted from the supercontinuum generated by tightly focusing the femtosecond laser pulses in water. The average velocity maps computed using time-correlated image-pairs and spray edge complexity computed using the average curvature scale space maps are compared for the spray images obtained with the two illumination techniques and also for the numerically simulated spray using the coupled volume of fluid and level set method for interface tracking (direct numerical simulation or DNS). The spray images obtained with supercontinuum-derived, incoherent, ultrafast illumination are clearer, since the artifacts arising due to laser speckles and multiple diffraction effects are largely reduced and show a better correlation with the DNS results.

**Introduction**
The enhancement in the efficiency and reduction in the emitted pollutants by the liquid or gaseous fuel based combustion engines requires a detailed understanding of various processes leading up to the combustion of the fuel, like injection, atomization, vaporization, etc. Among these processes fuel injection, leading to its atomization, plays a very crucial role since it is one of the initial steps and determines the drop size distribution, which in turn influences the liquid evaporation rate and the fuel-air mixture efficiency inside the combustion chamber and hence, the efficiency of the whole combustion process largely depends on it. To this end, several optical diagnostic tools, most being non-intrusive in nature, have been proposed to understand each of these processes individually [1]. The general trend in Diesel applications so far has been to increase the injection pressure (up to 2000 bars) and to decrease the orifice diameter (down to 100 μm), as these two parameters seem to improve the engine efficiency, but still the breakup of the spray is not fully understood. The main reason is that the form and distribution of liquid structures in the spray is a result of a lot of interdependent and complex processes such as hydrodynamic instability, cavitation, turbulence, etc. [2], which complicate the analysis of atomization phenomena and prevent direct control of the spray formation.

This work aims at improving the classical imaging methods particularly ultrafast shadow imaging for the high-pressure fuel sprays [3] by replacing the coherent ultrafast laser source by an incoherent ultrafast source, derived from the supercontinuum generated using a femtosecond laser. Since the light source used for imaging is now incoherent, the artifacts arising due to speckles and multiple diffraction effects are reduced. For quantitative comparison, interface velocity maps and complexity are computed from the time-correlated image-pairs and average curvature scale space (CSS) map respectively, extracted from a large set of images obtained using these two kinds of illumination sources and the results are compared with the simulated version of the spray computed using coupled volume of fluid and level set method for interface tracking, i.e. by direct numerical simulation (DNS). The incompressible Navier-Stokes equations are solved following a projection method and coupled to a transport equation for the level set function on a Cartesian meshing [4, 5].

**Material and methods**
The experimental setup for measuring the velocity of the spray in the near field of the nozzle based on the image-pairs is shown in figure 1 with two different illuminations (a) classical direct imaging using laser (coherent) source and (b) supercontinuum-derived, incoherent source. The speeds of the high-pressure fuel sprays probed in this work in the near field of the nozzle is about 350 m/s. Hence for imaging an ultra-short pulsed illumination source is required to freeze the motion of the spray. Here we use paired femtosecond laser pulses, each with a pulse duration of about a 100 fs and energy of 3.5 mJ per pulse, centered at wavelength $\lambda$ = 800 nm (FWHM = 12 nm), generated by two separate regenerative amplifiers (Coherent Libra) seeded by a common mode-locked Titanium-Sapphire oscillator (Coherent

Vitesse) at a repetition rate of 1 Hz (reduced from 1KHz, due to the restrictions imposed by the injector control). The time-delay between the two pulses was adjusted so that the displacement of the liquid structures is perceivable between the two images, while keeping almost the same shape. This delay, typically between 200 to 300 ns for the high-pressure injectors used in this study, is compatible with the double-frame CMOS camera (LaVision) for recording paired images.

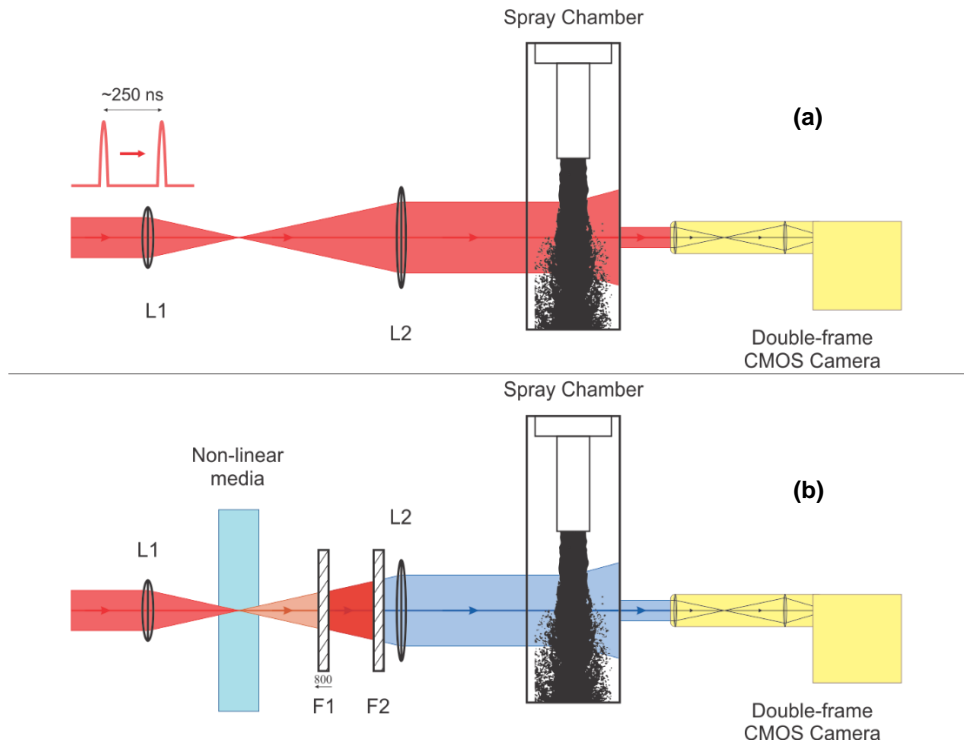

**Figure 1.** Schematics of the experimental setups for imaging of fuel sprays (a) with direct laser (coherent) illumination and (b) with supercontinuum-derived, incoherent illumination.

The ultrafast incoherent illumination source was extracted from the supercontinuum, generated by tightly focusing the high-power femtosecond laser pulses inside a 10 mm quartz cuvette filled with distilled water [6]. A small section of this supercontinuum centered at $\lambda = 450$ nm with FWHM = 40 nm was selected, using an appropriate band-pass filter (F2), for the illumination of the fuel sprays. Figure 2a shows the full spectra of the generated supercontinuum and figure 2b shows the selected wavelengths for the illumination of the spray.

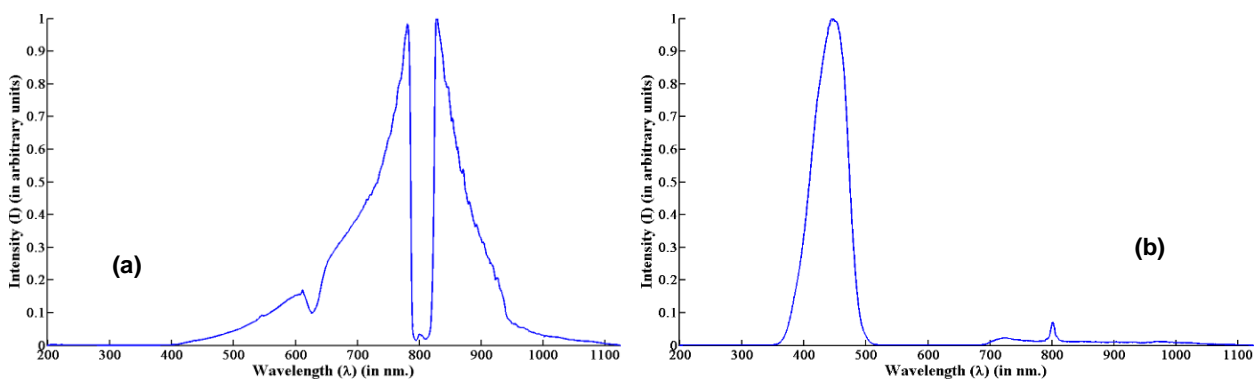

**Figure 2.** (a) Full spectrum for the supercontinuum, generated by focusing 800 nm fs laser pulses in water. (b) The selected bandwidth for illumination (or imaging) of the fuel sprays using a 450/40 nm band pass filter.

Note that a lot of non-linear processes, for example, self-phase modulation, filamentation, self-focusing, etc. are responsible for the supercontinuum generation and due to the combined effect of these processes there is a huge spectral broadening leading to a very short coherence lengths of the order of a few microns, whereas the pulse width or pulse duration is not changed significantly for individual wavelengths [7]. Nevertheless, the overall pulse width might increase due to dispersion while the beam travels through its path. Figure 3 shows the optical Kerr response of the 1.0 mm liquid carbon disulfide ($CS_2$) when illuminated by the generated supercontinuum. The abscissa shows the temporal delay between the pump (~100 fs, 800 nm) and the probe (supercontinuum pulse, overall pulse width certainly greater than 100 fs due to dispersion) pulses [8]. Figure 3 gives a broad idea about the overall pulse width for the band of wavelengths (410 – 490 nm) extracted from the supercontinuum. It should be noted that even after the dispersion the overall pulse width is small enough (a few picoseconds) to freeze the motion of the fuel sprays.

An alternative, Shell NormaFluid - ISO 4113, to diesel fuel with similar thermo-physical properties, listed in Table 1 [9], was used for the experiments presented in this work. This substitute liquid was injected inside a spray chamber maintained at ambient pressure and temperature using a single-orifice injector with diameter, Φ = 180 µm, with injection pressure of about 300 bars.

**Table 1:** Thermo-physical properties of the substitute liquid used in this work (ISO 4113).

| Density | Viscosity | Surface Tension |
|---|---|---|
| 821 kg.m$^{-3}$ | 0.0032 kg.m$^{-1}$.s$^{-1}$ | 0.02547 N.m$^{-1}$ |

Qualitatively, the images obtained with supercontinuum-derived incoherent illumination source are better than the images obtained directly with the laser. Figure 4 shows a magnified view of a small section of the spray image obtained by the two kinds of illuminations. It is clear that the structures are better defined and that speckles and multiple diffraction effects are highly reduced in the images obtained with supercontinuum derived illumination [10]. These images were further subjected to sophisticated image processing tools for quantitative comparison between them based on the interface velocity and complexity as described in the following section.

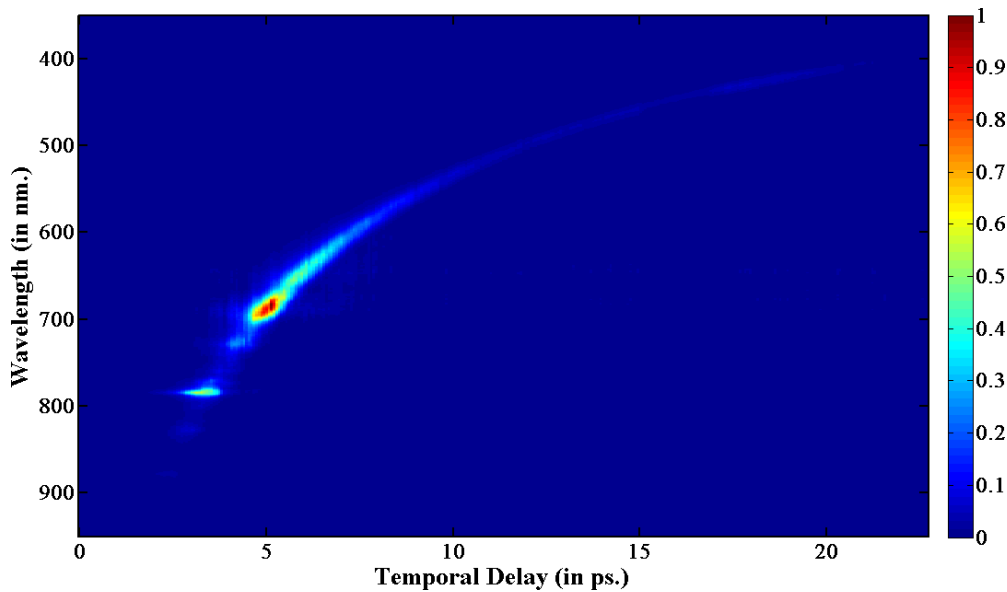

**Figure 3.** Supercontinuum chirp characteristics.

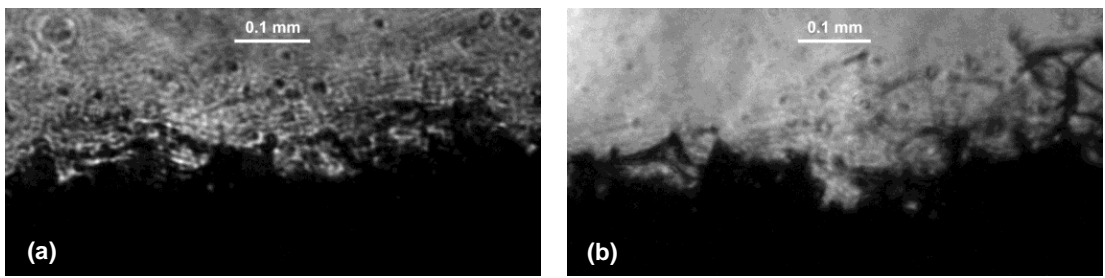

**Figure 4.** A magnified view of a section of the spray image through a single orifice nozzle with orifice diameter of 180 µm obtained with (a) direct laser light illumination and (b) supercontinuum-derived illumination.

**Quantitative comparison tools**
A large set of image-pairs recorded with direct laser and supercontinuum-derived illumination sources, so as to extract the statistically significant differences, were subjected to (i) cross-correlation based velocity computation method from the image-pairs and (ii) interface complexity calculations through curvature scale space computation (CSS) on the extracted spray edges/boundaries.

*Cross-correlation based velocity computation*
The computation of the velocity from the recorded image pairs consists in calculating the cross-correlation between the two paired images and determining the shift in the spray pattern [3]. A small section (square window) centered around a point near the edge of the spray is chosen from the first image of the pair and a larger window centered around the same point is chosen on the second image. The location of the maximum value in the two-dimensional normalized cross-correlation map of these two sections gives the displacement of the spray pattern. Figure 5 shows an example explaining the applicability of this technique to a spray image-pair.

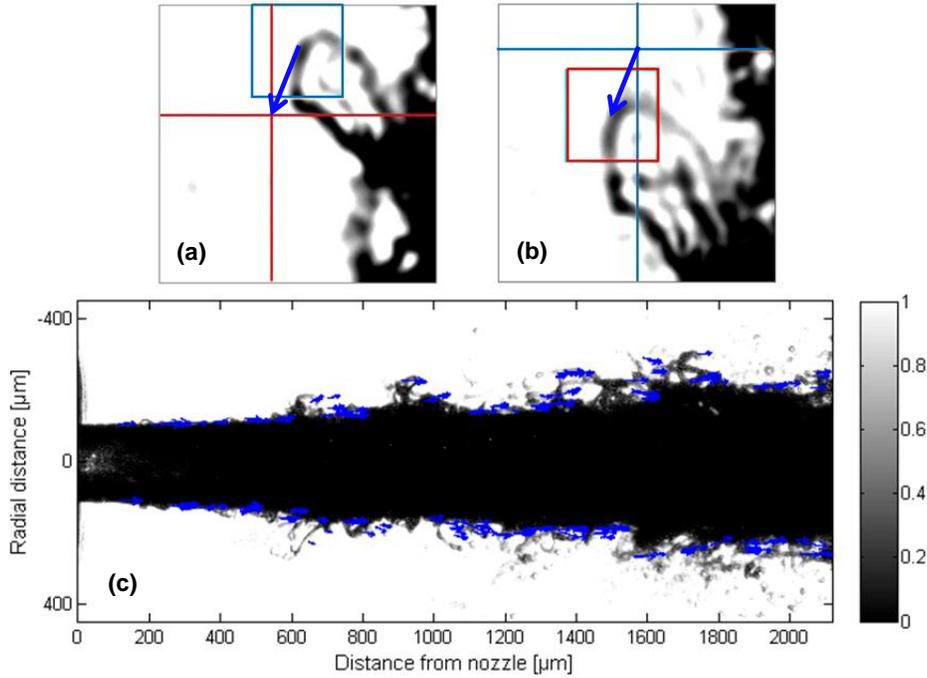

**Figure 5.** Images (a) and (b) shows the selected sections from time-correlated spray images-pairs and the corresponding displacement vector computed by correlating the two images. (c) Spray image with computed displacement vectors close to the spray edges.

Since the time between the two images is known a-priori, a velocity map can be deduced and averaged over the entire set of the recorded images. The results are shown in the following section.

*Curvature scale space computation*
The computation of curvature scale space (CSS) consists in extracting the edges or boundaries of the core jet using basic image processing tools, most of them included in the OpenCV image-processing library [11]. Let $x(t), y(t)$ denote the extracted edge of the spray with the parameterization variable $t$. The extracted spray edge or boundary is then smoothened using a Gaussian kernel of different widths $\sigma$ [12]. The curvature scale space $k(t,\sigma)$ is mathematically given by,

$$k(t,\sigma) = \frac{X'(t,\sigma)Y''(t,\sigma) - X''(t,\sigma)Y'(t,\sigma)}{\left(\left(X'(t,\sigma)\right)^2 + \left(Y'(t,\sigma)\right)^2\right)^{3/2}} \qquad (1)$$

where $X'$ and $X''$ indicate first and second derivatives of $X$ with respect to the curve parameterization variable $t$. $X(t,\sigma)$ and $Y(t,\sigma)$ are obtained by convoluting the extracted curve $x(t), y(t)$ with a Gaussian of width $\sigma$,

$$X(t,\sigma) = x(t) \otimes g(t,\sigma)$$
$$Y(t,\sigma) = y(t) \otimes g(t,\sigma)$$
(2)

where $\otimes$ denotes the convolution operation with,

$$g(t,\sigma) = \frac{1}{\sigma\sqrt{2\pi}}\exp(-t^2/2\sigma^2)$$
(3)

Note that as the value of $\sigma$ increases, the extracted edge smoothens out tending towards a straight line. Also, as one moves away from the nozzle in the direction of the flow, the complexity of the spray tends to increase. Figure 6a shows the original extracted spray boundary together with smoothened curves using the Gaussian kernel of different widths, $\sigma$ = 33 pixels and 129 pixels. Figure 6b shows a typical CSS map.

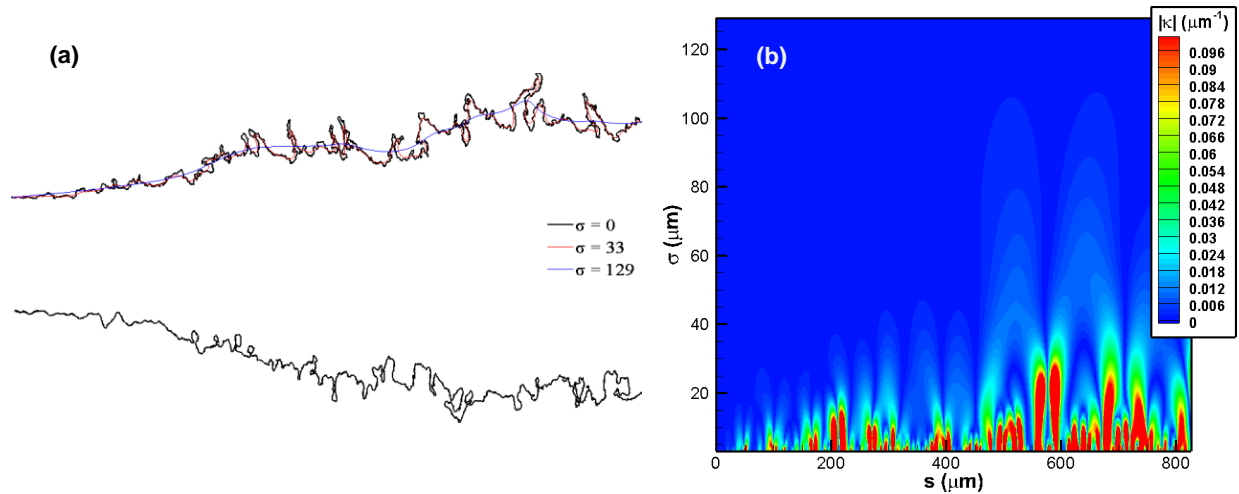

**Figure 6.** (a) The extracted boundary of the spray (in black, for σ = 0) and the smoothened boundaries using a Gaussian kernel of width, σ = 33 pixels and 129 pixels. (b) CSS map.

**Results and Discussion**

The velocity maps were constructed from a set of 500 time-correlated image-pairs to obtain a statistically significant result. Before subjecting these image-pairs for velocity computation, these images were normalized so as to remove the effect of the difference in the intensities of the two laser pulses and then for each image the core jet and the disconnected droplets were separated and then the velocity maps for the two were constructed separately for better understanding and visualization [13]. Figure 7 shows the average velocity maps for the core jet and disconnected droplets with direct laser light (coherent) illumination and supercontinuum-derived, incoherent illumination.

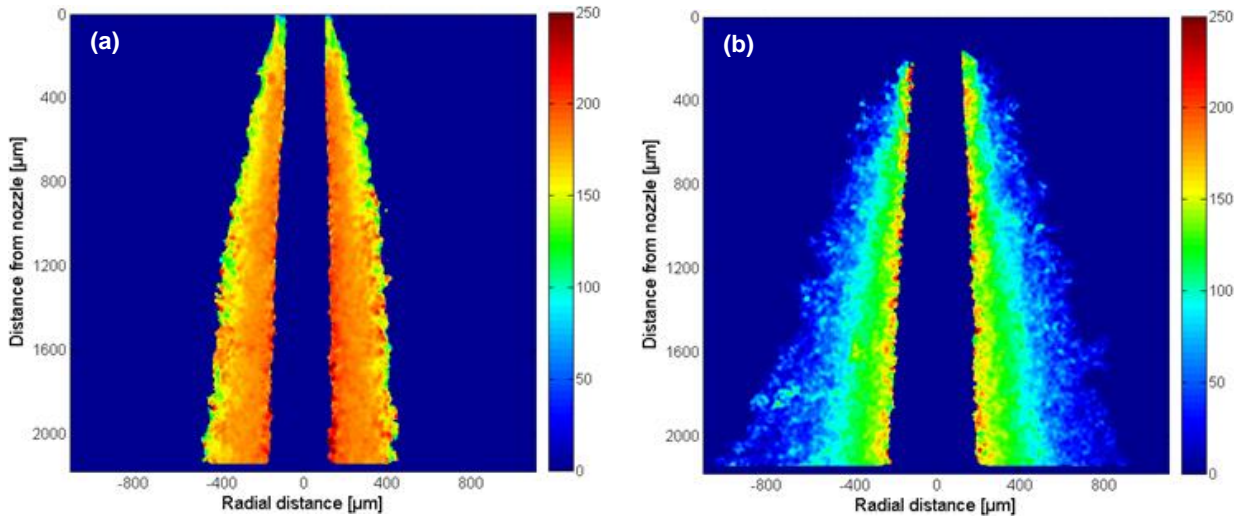

ILASS – Europe 2014, 8-10 Sep. 2014, Bremen, Germany

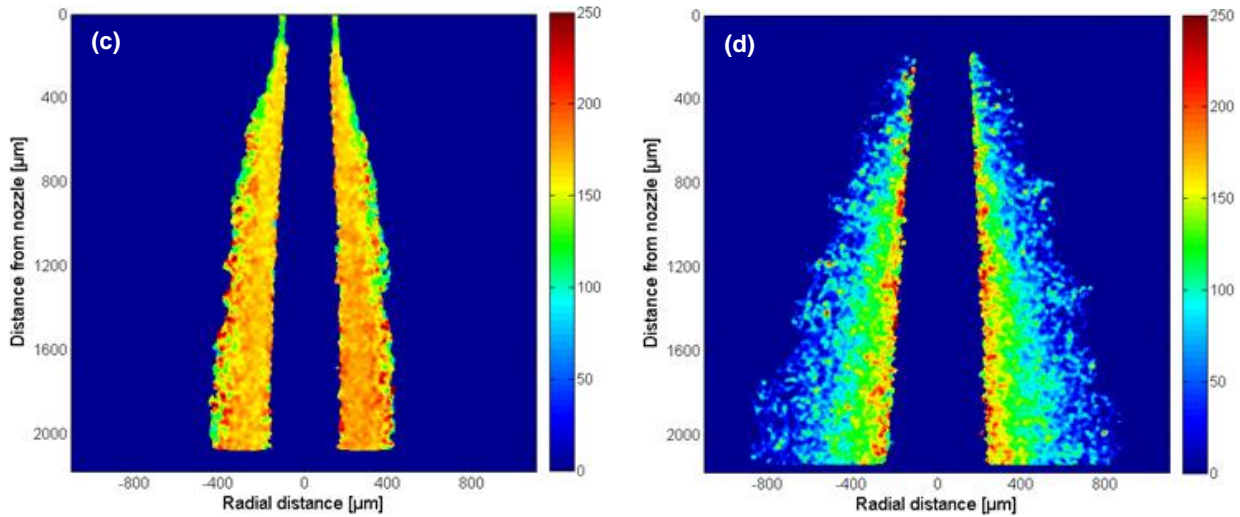

**Figure 7.** Average velocity maps (values in m/s) for (a) core jet, (b) disconnected droplets with direct laser (coherent), illumination, (c) core jet and (d) disconnected droplets with supercontinuum-derived, incoherent illumination.

As we expected, the average velocity maps for the core jet look very similar, whereas for the disconnected droplets even though the magnitudes of the velocities are not very different, some differences can be spotted near the edges in the average velocity maps with coherent and incoherent illuminations. The noise near the outer edge of the spray droplets arising due to speckles and multiple diffraction artifacts is minimal in case of the supercontinuum-derived, incoherent illumination.

The curvature scale space (as in eqn. (1)) was computed to quantify the complexity of the jet at different scales for the two cases of illumination. The scale was varied from 0 to $\sigma_m$, where $\sigma_m$ is the minimum value of scale for which the extracted edge of the spray resembles a straight line. The results were compared with the numerically simulated spray computed using the two-phase flow computation code (ARCHER [4]) based on coupled level set and volume of fluid method [14]. And again to extract the statistically significant differences, an average was taken over a large set of data (500 images). To minimize the minor fluctuations in the values of the curvatures the images were divided into small horizontal strips and an average value of the curvature was calculated and the CSS maps were constructed. Figure 8 shows the average CSS maps for the left and right (or bottom and top) edge of the spray with (a, b) direct laser illumination, (c, d) supercontinuum-derived illumination and (e, f) for the numerically simulated spray.

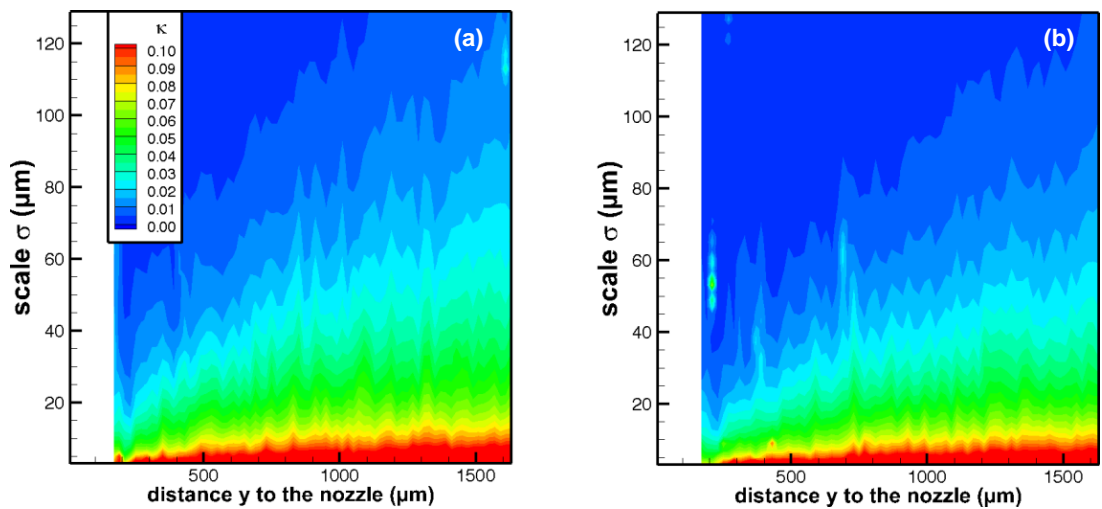

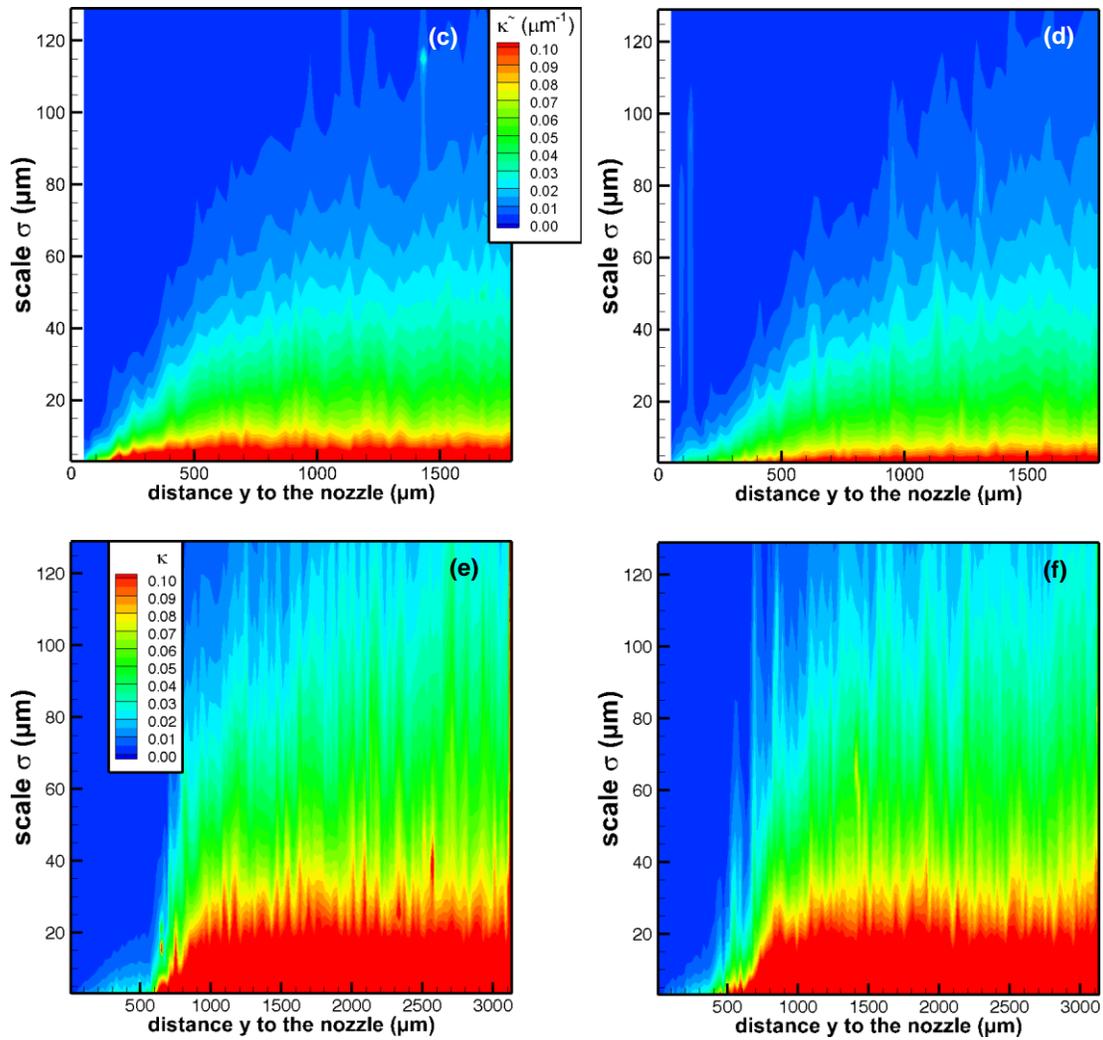

**Figure 8.** Average curvature scale space (CSS) maps (in µm$^{-1}$) for the left and right (or bottom and top) edge of the spray with (a, b) direct laser light illumination, (c, d) supercontinuum-derived, incoherent illumination and (e, f) numerically simulated spray.

Since it is difficult to reach to a definite conclusion directly from these average curvature maps, the variation of scale ($\sigma$) as a function of the distance from the nozzle ($y$) was obtained by multiplying all the curves for $\kappa = 0.01$ to $\kappa = 0.09$ of figure 8 by an appropriate factor to coincide with the curve for $\kappa = 0.1$. This was done for both the edges of the spray and the results are shown in figure 9.

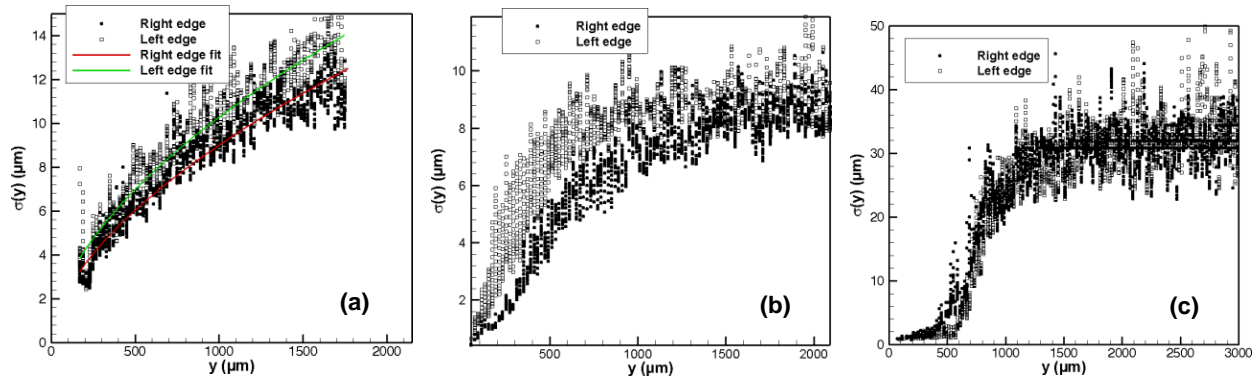

**Figure 9.** Variation of scale ($\sigma$) as a function of the distance from the nozzle ($y$) with (a) direct laser illumination, (b) supercontinuum-derived, incoherent illumination and (c) for the numerically simulated spray.

From figure 9 it seems that for the images obtained with the direct laser illumination the scale keeps on increasing with the distance $y$ from the nozzle whereas for the images obtained with supercontinuum-derived illumination the complexity saturates after a certain distance from the nozzle. A comparison with the numerical simulations (figure 9c) show that supercontinuum illuminated images are similar to the DNS calculations and then it is reasonable to claim that they give more accurate results. This can be explained by the fact that the edges are more clearly defined in the images obtained with this kind of illumination, being free from the artifacts of laser speckles and multiple diffraction patterns.

**Conclusions**

There is a strong demand for high-quality spray measurements to validate the numerical models of breakup in order to improve the fundamental understanding of the atomization processes. This proceeding presents a quantitative comparison between the two illumination sources – coherent (fs laser) and incoherent (derived from supercontinuum). High resolution time-correlated image-pairs were obtained by using a trans-illumination imaging arrangement using a double-pulsed femtosecond laser system coupled to a double frame CMOS camera. The sprays used in this work were produced by a single-hole, plain orifice injector assembly dispersing fuel oil into ambient atmospheric conditions. The high resolution images allowed us to isolate the dispersed liquid elements from the jet core, facilitating a more refined examination of the correlated image structure velocities. Direct numerical simulations of a high-pressure liquid jet have been carried out using a coupled volume of fluid/level set (VOF/LS) method for interface tracking.

The average velocity maps obtained for the two illuminations appear to be very similar although there were certain subtle differences near the spray boundaries. The average CSS maps were used to compare the interface complexity. A very different behavior is observed for the two kinds of illuminations. It has been shown that the images obtained with supercontinuum-derived, incoherent illumination resemble the images obtained using numerical simulations.